\documentclass[11pt]{article}
\usepackage{amsmath}
\usepackage{amsfonts}
\usepackage{color}

\title{Unitary Representations of Quantum Superpositions of two Coherent States and beyond}
\author{Antonino Messina\\{\footnotesize\it Dipartimento di Fisica e Chimica,
Universit\` a degli Studi di Palermo,}\\{\footnotesize\it Via Archirafi 36, 90123 Palermo, Italy
 \& antonino.messina@unipa.it}\\[2ex]
Gheorghe Dr\u ag\u anescu\\{\footnotesize\it Faculty of Mechanics, Polytechnic
 University of Timi\c soara,}\\{\footnotesize\it
Bd. M. Viteazu, Nr. 1, 300222, Timi\c soara, Romania \& ghed@mec.upt.ro}}

\begin{document}

\maketitle
\begin{abstract}
The construction of a class of unitary operators generating linear superpositions of generalized coherent states from the ground state of a quantum harmonic oscillator is reported. Such a construction, based on the properties of a new ad hoc introduced set of hermitian operators, leads to the definition of new basis in the oscillator Hilbert space, extending in a natural way the displaced Fock states basis. The potential development of our method and our results is briefly outlined.
\end{abstract}

\section{Introduction}

The coherent states of a quantum harmonic oscillator are minimum uncertainty
 non stationary states often regarded as quasi classical states since, as
  Schr\" odinger firstly showed in 1926 \cite{1},
the dynamical evolution of the canonical variables vividly resembles that
 of the corresponding
variables of a classically evolving oscillator. Many years later Glauber
 published some seminal
papers \cite{2} where he demonstrated the relevance of such states for the quantum treatment of
optical coherence \cite{3}. Since then, the coherent states have been theoretically and experimentally
investigated initially in quantum optics mainly \cite{4,5} and subsequently extended in different
mathematical \cite{6,7,8,9} and physical \cite{10} contexts like, for example, trapped ions
\cite{10,11,12} and quantum information processing \cite{13,14,15}. The recently published special issue
\cite{16} on the research activity on the mathematical and physical facets of the coherent states
provides an excellent glance on the related state of the art.

As coherent states manifest a weak "quantumness", particular attention has been devoted to the
properties exhibited by superpositions of coherent states since such states provide the very
attractive occasion to highlight bizarre and counterintuitive features of quantum behaviour.
Quantum superpositions of coherent states possessing equal high amplitudes but phases differing by $\pi$
may be viewed for instance as macroscopically distinguishable quantum superpositions \cite{17,18,19,20}
whose robustness against environmental decoherence has attracted many experimentalists, for example in
cavity quantum electrodynamics, in connection with the quantum to classical transition \cite{21,22}.

It is well known that the displacement operator unitarily generates coherent states from the ground state
of the oscillator and most of the advantages stemming from such a representation have been fully
brought to light. It seems therefore reasonable to figure out that knowing how to generate from the
oscillator vacuum state special classes of generalized coherent states with the help of unitary
operators appropriately generalizing the displacement operator, might provide an additional tool
to disclose unusual properties exhibited by such states.

In this paper we report the mathematical construction of some unitary representations of quantum
superpositions of two generalized coherent states. Such a construction is driven
by the spectral properties of new and ad hoc introduced hermitian operators. Then, in analogy with
the displacement operator which acting upon the Fock number state basis generates the new basis
called Displaced Fock States \cite{23}, today recognized of relevance in many areas and in particular
in quantum optics, the knowledge of such unitary operators is here exploited to define new basis
of quantum states where the harmonic oscillator should exhibit non-classical properties worthy to
be brought to light and elucidated.

\section{The parity-displacement operator}

The stationary Fock states $\{ |n>, n \in N \}$ of a quantum harmonic oscillator
generate the Fock space where the two ladder operators $a$ and $a^\dag$ satisfy
the following properties:

\begin{equation}
a | n > = \sqrt{n} |n-1>, \quad  a^\dag | n+1 > = \sqrt{n+1} |n+1>, \quad
a^\dag a | n > = n |n>,
\label{1}
\end{equation}
and $[ a, a^\dag ] = 1$. The operator $\cos \pi a^\dag a= \exp(i \pi a^\dag
 a)$ introduced by Glauber in 1963 is called parity operator since $\cos
\pi a^\dag a |n> = (-1)^n |n>$. It is unitary, hermitian, and anticommutes
with $a$ and $a^\dag$. The Glauber normalized coherent state of a complex
dimensionless amplitude $z$, defined in the Fock space as:

\begin{equation}
|z> = \exp(-\frac{|z|^2}{2}) \sum_{n=0}^\infty \frac{z^n}{n} |n>,
\label{2}
\end{equation}
is either eigenstate of the annihilation operator $a$ of eigenvalue $z$,
$a|z> = z |z>$, or the displacement of the ground state of the harmonic oscillator
in its phase space accomplished by the so called unitary displacement operator

\begin{equation}
D(z) = \exp(z a^\dag - z^\ast a), \Rightarrow D(z)  |0> = |z>.
\label{3}
\end{equation}
satisfying the properties:

\begin{equation}
D(z) D(z') = \exp(i \mbox{Im}(z z'^\ast)) D(z + z'), \qquad \mbox{and} \qquad
D^\dag(z)  = D(-z).
\label{4}
\end{equation}

Since $a \cos( \pi a^\dag a) = - \cos( \pi a^\dag a) a$, then $ \cos( \pi a^\dag a)|z> =|-z>$, so that

\begin{equation}
B(z) = D(z) \cos(\pi a^\dag a) = \cos(\pi a^\dag a) D(-z) =  [\cos(\pi a^\dag
a)]^\dag D^\dag (z)  = B^\dag (z).
\label{5}
\end{equation}
that is the operator $B(z)$  is hermitian, besides being unitary as product
 of two unitary operators. We call this operator the \textit{parity-displacement
 operator}. It provides the tool for the first step of our construction
of an unitary  representation of a quantum superposition of two coherent
states. $B(z)$ has only two possible eigenvalues +1  and -1, each of them
of infinite degeneracy. The main property of $B(z)$  making it of interest
for the scope of this paper is simply its action upon the coherent state
$|z>$:

\begin{equation}
B(z) |z> = D(z) \cos(\pi a^\dag a) |z>  = D(z) |-z>  =  D(z) D(-z) |0>  =
|0>,
\label{6}
\end{equation}
since $\mbox{Im}(z (-z^\ast)) = 0$. Considering that, on the other hand, $B(z)|0>
 =|z>$ we are legitimated to search eigenstates of $B(z)$ in the bidimensional subspace
 spanned by the two nonorthogonal coherent vectors $|0>$ and $|z>$. Solving
the two equations $B(z)(C_0^\pm |0> + C_2^\pm |0>)= \pm (C_0^\pm |z> + C_2^\pm
|z>)$ we achieve the following two orthogonal normalized eigenvectors of
$B(z)$:

\begin{equation}
|b_\pm (z)> = N_\pm^{-1} (|0> \pm |z>),
\label{7}
\end{equation}
with $N_\pm = [2 (1 \pm e^{-\frac{1}{2}|z|^2})]^{1/2}$ since $<0|z> = <z|0>
= e^{-\frac{1}{2}|z|^2}$ in view of eq. (\ref{2}). It is thus easy to express
the ground state of the oscillator as quantum superpositions of the states
$|b_+ (z)>$ and $|b_-(z)>$ as follows:

\begin{equation}
|0> = \frac{1}{2}[N_+ |b_+ (z)> + N_- |b_- (z)>],
\label{8}
\end{equation}

An immediate consequence of eqn. (\ref{8}) is that the action upon $|0>$
of the unitary $\lambda$ evolution-like operator $\exp(i \lambda B(z))$, $\lambda \in R$, can be easily evaluated and cast in the form:

\begin{equation}
\exp(i \lambda B(z)) |0> = \frac{N_+}{2} e^{i \lambda}|b_+> + \frac{N_-}{2}
e^{-i \lambda} |b_-> ,
\label{9}
\end{equation}
which, in turn, may be expressed as linear combination of the two coherent
states $|0>$ and $|z>$ as follows:

\begin{equation}
U(\lambda; z)|0> = \exp(i \lambda B(z)) |0> = \cos \lambda |0> + i \sin \lambda
 |z> = |B(\lambda)>,
\label{10}
\end{equation}

Thus like the coherent state $|z>$ arises from the vacuum state under the
action of the unitary displacement operator $D(z)$, the class of the normalized
$\lambda$-parametrised quantum superposition $|B(\lambda)>$ of the two prefixed
coherent states $|0>$ and $|z>$ arises as well from the vacuum state under
the action of the $\lambda$ evolution-like unitary operator $U(\lambda,z)$
associated to the parity-displacement operator $B(z)$. We underline that
since $<0|z> \not= 0$, if we substituted in (\ref{10}) the quantum relative
phase $i = e^{i \frac{\pi}{2}}$ with $e^{i \phi}$, ($\phi \not= \pm \pi/2$),
the corresponding state would become unnormalized.

We are aware that $B(z)$ possesses infinitely-many bidimensional invariant
subspaces and then that eq. (7) only describes the pair of eigenstates of
$B(z)$ involving the vacuum state $|0>$ in accordance with the aim of this
paper. The consequences of this property of $B(z)$ will not be addressed
in this paper.

\section{Generalizing the parity-displacement operator}

Eq. (\ref{10}) expressing the action of $U(\lambda, z)$ on the state $|0>$
may be generalized as follows. Denoting by $m$ any positive integer, the complex number $e^{i\frac{\pi}{m}}=
e^{i\frac{2\pi}{2 m}}$ is a $m$-th root of -1 and a $2 m$-th primitive root
of +1. It is well known that the unitary operator $e^{i \frac{\pi}{m} a^\dag
a}$ transforms the annihilation operator as follows:

\begin{equation}
e^{-i \frac{\pi}{m} a^\dag a}a e^{i \frac{\pi}{m} a^\dag a} = e^{i\frac{\pi}{m}}
a
\label{11a}
\end{equation}
that implies that both  $e^{i \frac{\pi}{m} a^\dag a}$ and $e^{-i \frac{\pi}{m}
 a^\dag a}$ anticommutes (commutes) with the operator $a^m$ ($a^{2m}$). Since

\begin{equation}
\cos \tfrac{\pi}{m} a^\dag a = \frac{1}{2}(e^{i \frac{\pi}{m} a^\dag a}+ e^{-i
 \frac{\pi}{m} a^\dag a})
\label{12a}
\end{equation}
we immediately get $[\cos \frac{\pi}{m} a^\dag a, a^{2m}] = 0$ and:

\begin{equation}
\{ \cos \tfrac{\pi}{m} a^\dag a, a^{2 m} \} =  \cos(\tfrac{\pi}{m} a^\dag a)
 a^{2 m} +  a^{2 m} \cos(\tfrac{\pi}{m} a^\dag a)  = 0,
\label{13a}
\end{equation}
whatever the positive integer $m$ is. Thus the anticommutation between the
parity operator $\cos\pi a^\dag a$ and $a$ appears to be a special case of
the property expressed by eq. (\ref{13a}).
It is worth noticing that differently from $\cos \pi a^\dag a$, the operator
$\cos \frac{\pi}{m} a^\dag a$  is not unitary when $m>1$, while of course,
is still hermitian for any $m \ge 1$, its $m+1$ eigenvalues being:
 1, $\cos(\frac{\pi}{m})$, ... $\cos(\frac{n \pi}{m})$, ...
$\cos(\frac{(m-1)\pi}{m})$, -1.

Eq. (\ref{13a}) suggests a generalisation  of the parity displacement operator
$B(z)$ defined by eq. (\ref{5}). Let's introduce the $m$ parity-displacement
 operator $B_m (z)$:

\begin{equation}
B_m (z) = \exp\{\frac{(-1)^m}{m} (z^\ast a^m - z (a^\dag)^m)\} \cos \tfrac{\pi}{m} a^\dag a = D_m (z) \cos \tfrac{\pi}{m} a^\dag a,
\label{14a}
\end{equation}
which for $m = 1$ traces back to the operator $B(z)$ sharing with it the property of being hermitian. Indeed:

\begin{equation}
B_m (z) =  \cos(\tfrac{\pi}{m} a^\dag a) D_m (-z) =  (\cos \tfrac{\pi}{m} a^\dag
 a)^\dag D_m^\dag (z) = B_m^\dag (z).
\label{15a}
\end{equation}

It is immediate to identify $D_2(z)$ as a squeeze operator of squeeze parameter
$|z|$, generally denoted by $S(z)$ in the literature. At the best of our
knowledge, the properties of the unitary operators $D_m (z)$ with $m>2$ have
not been investigated in detail so far and since no special name has ever
been assigned to it, we call $D_m(z)$ the $m$ {\em generalized displacement
operator}. Formally we introduce the following notation for the action of
 $D_m (z)$ on the vacuum state $|0>$:

\begin{equation}
D_m (z) | 0 > = | z_m >.
\label{16a}
\end{equation}

The state $|z_1>$ coincides with the coherent state $|z>$
while $| z_2 >$ coincides with the squeezed vacuum state $S(z) |0> = D(0)
 S(z)$ $|0> = |(0,z)> = S(z) D(0) |0> = |[z,0]>$, where the (coincident)
 states $|(0,z)>$ and $|[z,0]>$ are special cases of the Caves ideal squeezed
  states $|(v,z)>$ $ = D(v)S(z)|0>$ and of the Yuen two photon coherent state $|[z,v]>= D(v)S(z)$ $|0>$ \cite{24}.

Adopting the same convention used in the literature for $|z_1>$  and $|z_2>$ we fix the global phase of $|z_m>$ setting $<0|D_m(z)|0>$  real.

 In passing we note that even if we do not know the number statistics of
$|z_m>$ for $m>2$, it is however possible to show that the operators
$\cos \frac{\pi}{m} a^\dag a$ and $\sin \frac{\pi}{m} a^\dag a$ acting
on $|z_m>$ fulfill properties coincident with those possessed by $\cos
\pi a^\dag a$  and $\sin \pi a^\dag a$ when applied on a coherent state $|z>$.
 Indeed:

\begin{equation}
\cos \tfrac{\pi}{m} \pi a^\dag a |z_m> = \cos[\tfrac{\pi}{m} \pi a^\dag a]
 D_m(z) |0> = |(-z)_m>
\end{equation}

\begin{equation}
\sin[\tfrac{\pi}{m} \pi a^\dag a] |z_m> = D_m (-z) \sin[\tfrac{\pi}{m} \pi a^\dag a] |0> = 0.
\end{equation}

It is remarkable that $B_m (z)$ shares with $B(z)$
properties exploitable to get a result that generalizes eq. (\ref{10}). To
 appreciate this point we first apply $B_m (z)$  on the vacuum state  $|0>$,
formally obtaining

\begin{equation}
B_m |0> = D_m(z) \cos(\tfrac{\pi}{m} a^\dag a) | 0 > = D_m (z) |0> = |z_m>.
\label{17a}
\end{equation}

On the other hand applying $B(z)$ to $|z_m>$ yields

$$
B_m (z) |z_m> = D_m(z) \cos(\tfrac{\pi}{m} a^\dag a) |z_m> = \cos(\tfrac{\pi}{m}
a^\dag a)D_m(-z) D_m(z) |0> =
$$

\begin{equation}
= \cos(\tfrac{\pi}{m} a^\dag a)D_m(z) D_m^\dag(z) |0> = \cos(\tfrac{\pi}{m}
a^\dag a) |0> = |0>.
\label{18a}
\end{equation}
for any $m \ge 1$. Taking into account that the operator $\sin\frac{\pi}{(m)}a^\dag
a$ is diagonal on the Fock states basis, eq. (\ref{18a}) implies that $<0|z_m>=0$
as soon as $n \not= k m$ with $k = 1, 2, ...$ Thus $|z_m>$ contains only
 number states multiples of $m$. This property, explicitely verifiable for
 any squeezed vacuum state, suggests to call the state  $|z_m>$: {\em m  multiple generalized coherent state}.

Now by virtue of what already done going from eq. (\ref{6}) to
eq. (\ref{10}), we easily figure out the successful construction of a unitary operator leading as $U(\lambda,z)$ to representations of quantum superpositions of $|0>$ and $|z_m>$ whatever $m \ge 1$ is.  To this end we first notice that for $m >1$, $\cos \frac{\pi}{m}a^\dag a$ is not
unitary. Thus to look for orthogonal eigenvectors of $B_m (z)$ in the bidimensional
vector space spanned by $|0>$  and $|z_m>$, we must solve the equation $B_m(z)(c^{\lambda_m}_{0}
|0> + c^{\lambda_m}_{z_m} |z_m>) = \lambda_m (c^{\lambda_m}_0 |0> + c^{\lambda_m}_{z_m}
|z_m>)$. Since it gives rise to a $m$-independent linear and homogenous system
in the unknowns $c_0^{\lambda_m}$ and $c_z^{\lambda_m}$, the two normalized
(and then $m$-dependent) orthogonal eigensolutions of $B_m (z)$ we are searching,
in view of eq. (\ref{17a}) can be written down as follows:

\begin{equation}
|b_\pm^{(m)} > = [N_\pm^{(m)}]^{-1} (|0> \pm |z_m>)
\label{19a}
\end{equation}
belonging to the eigenvalues $\lambda_m = \pm 1$ respectively for any $m$.
The normalization constants $N_\pm$ have the form:

\begin{equation}
N_\pm^{(m)} = \sqrt{2 (1 \pm <0|z_m>)},
\label{20a}
\end{equation}
since the probability amplitude of $|0>$ in $|z_m>$ is real by definition.
Thus as in the case $m=1$

$$
|0> = \frac{1}{2} \left[N_+^{(m)} |b_+^{(m)}> + N_-^{(m)} |b_-^{(m)} > \right]
$$
leads to the easy definition of the  $\lambda$ evolution-like unitary operator:

\begin{equation}
U_m (\lambda;z) = \exp(i \lambda B_m (z)), \qquad \lambda \in R,
\label{21a}
\end{equation}
having the property:

\begin{equation}
U_2 (\lambda;z) |0> = \cos \lambda |0> + i \sin \lambda |z_m> \equiv |B_m(\lambda)>
\label{22a}
\end{equation}

For $m=1$ eq. (\ref{22a}) traces back to (\ref{10}) while for $m=2$, the
unitary operator $U_2 (\lambda;z)$ generates from the ground state a quantum
superposition between the same vacuum state and the state $|z_2> \equiv
|(0,z)> = S(z) |0>$ that is the squeezed vacuum state of the squeeze parameter
$|z|$.

It is of relevance that the action of $U_m (\lambda;z)$ on $|0>$ produces
a quantum superposition whose structure is $m$-independent.

In the next section we concentrate on such unitary operator to express it
 as closed linear function of $D_m(z)$ with operator coefficients diagonal
 in the Fock states basis .

\section{A linear expression of $U_m(\lambda;z)$}

By definition $U_m(\lambda;z)$ may be expanded as follows:

\begin{equation}
U_m(\lambda;z) = \sum_{k=0}^\infty \frac{(i \lambda)^k}{k!} B_m^k (z)
\label{23a}
\end{equation}

Observing that in view of eq. (\ref{13a})

\begin{equation}
B_m^2 (z) = [D_m(z) \cos(\tfrac{\pi}{m} a^\dag a)][D_m(z) \cos(\tfrac{\pi}{m}
 a^\dag a)] = \cos^2(\tfrac{\pi}{m} a^\dag a),
\label{24a}
\end{equation}
then, whatever the integer $k$ is, we get:

\begin{equation}
B_m^{2 k} (z) = \cos^{2 k}(\tfrac{\pi}{m} a^\dag a).
\label{25a}
\end{equation}

Thus the explicit expression of $U_m(\lambda;z)$, easily applicable  to a
generic Fock state $|n>$, may be cast as follows:

\begin{equation}
U_m(\lambda;z) = \cos[\lambda\cos(\tfrac{\pi}{m} a^\dag a) ] + iD_m(z) \sin[\lambda
\cos(\tfrac{\pi}{m} a^\dag a)]
\label{26a}
\end{equation}
which generates $|B_m(z)>$, as given by eq. (\ref{22a}) when applied to $|0>$.
It is possible to check that:

\begin{equation}
U_1 (\lambda;z) = \cos \lambda \mathbb{I} + i \sin \lambda D(z) \cos(\pi
a^\dag a) \equiv U(\lambda;z)
\label{27a}
\end{equation}
and

\begin{equation}
U_2 (\lambda;z) = \sin^2(\tfrac{\pi}{2} a^\dag a)\mathbb{I} + \cos \lambda \cos^2(\tfrac{\pi}{2}
a^\dag a) + i \sin \lambda D_2(z) \cos(\tfrac{\pi}{2}
a^\dag a),
\label{28a}
\end{equation}
which acting on $|0>$, give a quantum superposition between $|0>$ and the
coherent state $|z>$ or the squeezed vacuum state $S(z) |0>$ respectively.

By definition the operator $U_m(\lambda;z)$ satisfy the composition property:

\begin{equation}
U_m(\lambda;z) U_m(\lambda';z) = U_m(\lambda+\lambda';z),
\label{29a}
\end{equation}
which is easily checked using eq. (\ref{26a}) too.

We now exploit eq. (\ref{26a}) to construct a unitary representation of
linear combinations of two arbitrary coherent states $|z>$ and $|u>$. To
this end we introduce the following unitary operator:

\begin{equation}
V_1(\lambda;z,u) = D(\frac{u}{2}) U_1(\lambda;z) D(\frac{u}{2})
\label{30a}
\end{equation}
which gives:

\begin{equation}
V_1(\lambda;z,u) = \cos \lambda D(u) + i \sin\lambda e^{i \Im (u z^\ast)} B(z).
\label{31a}
\end{equation}
immediately yielding

\begin{equation}
V_1(\lambda;z,u) |0>  = \cos \lambda |u> + i \sin\lambda e^{i \Im (u z^\ast)} |z> \equiv |\Psi_0 (\lambda;z)>.
\label{32a}
\end{equation}

Choosing $u=-z$ implies $\Im(u z^\ast)=0$ and for $\lambda = \frac{\pi}{4}$
we get:

\begin{equation}
V_1(\frac{\pi}{4};z,-z) |0>  = \frac{1}{2} [|-z> + i |z>] = |\Psi_0 (\frac{\pi}{4};z)>,
\label{33a}
\end{equation}
 which is a special case of an equally weighted superposition of two opposite
 coherent states on a circle since of equal complex amplitudes and $\pi$
 difference of phase with each other.

Thus eq. (\ref{32a}) contains examples of those states termed cat-like states
introduced in the literature by Yurke and Stoler \cite{25} whose detection
 is of relevance in view of the fact that such states represent for $|z|$ large enough quantum
superpositions of two macroscopically distinguishable and essentially classical
states. An important difference between eq. (\ref{31a}) and eq. (\ref{27a})
is that the former does not possess a $\lambda$-evolutive structure by definition.

However the approach followed to get eq. (34) may be of some help to tackle
theoretical problems involving such kinds of superpositions.

To appreciate this point we wonder whether a result analogous to that expressed by eq. (\ref{32a}) may be reached for any  $m$. To this end we generalize
eq. (\ref{30a}) defining:

\begin{equation}
V_m (\lambda;z;u) =  D_m (\frac{u}{2}) U_m(\lambda;z)  D_m (\frac{u}{2}).
\label{34a}
\end{equation}
which taking into account that $\sin(\frac{\pi}{m} a^\dag a)$ too anticommutes
with $a$, may be expressed as:

\begin{equation}
V_m (\lambda;z,u) =  D_m^2(\frac{u}{2}) [\sin^2(\tfrac{\pi}{m} a^\dag a) + \cos \lambda
\cos^2(\tfrac{\pi}{m} a^\dag a)] + i \sin \lambda \tilde{D}_m (z)
\cos(\tfrac{\pi}{m} a^\dag a),
\label{35a}
\end{equation}
where $\tilde{D}_m (z)= D_m (\frac{u}{2})D_m(z)D_m (-\frac{u}{2})$. Applying
$V_m (\lambda;z,u)$ to $|0>$ we obtain:

\begin{equation}
V_m (\lambda;z,u) |0> = \cos \lambda D_m(\frac{u}{2}) |(\frac{u}{2})_m> +
i \sin \lambda D_m (\frac{u}{2}) D_m(z) |(\frac{u}{2})_m>.
\label{36a}
\end{equation}

Unfortunately eq. (\ref{36a}) could not be further simplified since we have
at our disposal no closed formula for the product of two $m$  generalized
displacement operators not at all easy to find out. However eq. (\ref{36a})
has a promising structure as well, since, given that for $m=2$   it provides
examples of quantum superpositions of two states in general macroscopically
distinguishable and manifesting squeezing \cite{14}, it opens up the possibility
that the cases with larger values of $m$  might hide nonclassical facets
beyond squeezing.

\section{New basis in the Fock space}

The main scope of this section is to exploit eq.(\ref{26a}) to define new
m-dependent basis through the action of $U_m (\lambda; z)$ on the number
states basis. This approach resembles the one leading from the number basis
to the displaced Fock states $\{|\alpha,n>\}$ with the help of the displacement
 operator $D(z)$. These non classical states have been quite recently
 experimentally simulated in a photonic lattice of evanescently coupled waveguides
\cite{14a}.

Our starting point is eq. (\ref{26a}) expressing the operator $U_m(\lambda;z)$
 in a form convenient for its application on a generic number state $|n>$.

Let's introduce the the symbol $|(\lambda;z_m), n>$ to denote the result of
 the application of $U_m(\lambda;z)$ on $|n>$, that is:

\begin{equation}
|(\lambda;z_m), n> =U_m(\lambda;z) |n>, \qquad n = 0, 1, ...
\label{37a}
\end{equation}

From eq. (\ref{26a}) we get:

\begin{equation}
|(\lambda;z_m), n> = \cos[\lambda \cos\tfrac{n}{m} \pi] |n> + i \sin[\lambda
 \cos\tfrac{n}{m} \pi] D_m(z) |n>
\label{38a}
\end{equation}

Eq. (\ref{38a}) describes the $(n+1)$-th state to the  $m$-labeled new
($m>2$) basis as linear combination of two $(n+1)$-th states of two orthonormal
 basis, namely the Fock basis and the basis $[ D_m(z) |n> \equiv |z_m,n>,
\, n=0, 1, 2... ]$ which extends in very transparent way the displaced Fock states basis $[ D (z) |n> \, n=0, 1, 2... ]$. Thus while acting on the vacuum   $U_m(\lambda;z)$ generates a linear combination of $|0>$ and $|z_m> = D_m(z) |0> = |z_m, 0>$, when it acts on $|n>$ we get a $n$-dependent weighted linear combination of the Fock state $|n>$ and of the state $| z_m,n >$. We observe that for
$m=1$ $|z_1, n>$ gives the $(n+1)$-the displaced Fock state, and for $m=2$
$|z_2, n>$ is the squeezed $n$-bosons state that is the squeezing-transformed
Fock state $|n>$ \cite{15}. Eq. (\ref{38a}) means that the number statistics
of $|z_m, n>$ coincides with that of $|(\lambda,z_m), n>$ except for the
 probabily amplitude of the number state $|n>$. It is possible to demonstrate
  that $< (\lambda,z_m), n |(\lambda,z_m), n> = <z_m,n|z_m,n> = 1$ identically
 with respect to all parameters, implies that $<n|D_m(z)|n>$  is real for any natural $n$ and $m$, a well known result when $m=1$.

Representing indeed the normalized state $D_m(z)|n>$ as $|z_m,n> = \sum_{j=0}^\infty
$ $c_{m,n}^j |j>$ we get:

\begin{equation}
\sum_{j \not= n} |c_{m,n}^j|^2 \sin^2 \lambda + |<n|(\lambda,z_m),n>|^2 =
 1,
\label{39a}
\end{equation}
which implies $\mbox{Im}(<z|D_m(z)|n>) = 0$.

Since the explicit expressions of the number statistics of $|z_m,n>$ is not
known when $m>1$, our simple conclusion seems not reported in the
literature at the best of our knowledge.

\section{Conclusive remarks}

The new idea reported in this paper is the construction of the $m$-labeled
unitary operators $U_m(\lambda,z)$ based on the introduction of the set of the generalized parity-displacement hermitian operators $B_m(z)$. By definition
$U_m(\lambda,z)$ possesses semigroup properties for any fixed value of  $m$ and $z$  and its  $\lambda$-evolutive nature generates from the vacuum state of the quantum harmonic oscillator a class of linear combinations of the same ground state $|0>$ and of the state $|z_m>$ resulting from the application
to $|0>$ of the $m$ generalized displacement operator $D_m(z)$ defined by
eq (\ref{14a}). This is our main result together with the derivation of a
closed fruitful expression of $U_m(\lambda,z)$ which enables us the
introduction of new $m$-labeled basis in the oscillator Hilbert space.
We believe that in view of the well studied nonclassical properties of the
displaced Fock states basis, in our notation $\{|z_1,n>\}$,  the  $m$-dependence
of the properties characterizing the basis $\{|z_m,n>\}$  are worthy to
be investigated with the aim of disclosing an  $m$-dependent quantumness
of these states to be compared with that exhibited by the displaced Fock states and by the squeezed $n$-bosons states, in our notation  $\{|z_2,n>\}$.

We conclude emphasizing that our method as well as our results, on the one hand, may be exported in a multimode scenario where the occurrence of multipartite entanglement is certainly an added stimulus to a systematic investigation of the emerging new states. On the other hand our approach puts indeed at our disposal a flexible enough tool to generate unitary representations of quantum superpositions of more than two coherent or   generalized coherent states of different controllably complex amplitude.

\section*{Acknowledgements}

AM express his gratitude to Riccardo Messina for stimulating conversation
on the subject of this paper.

\end{document}